\documentclass[a4paper,fleqn,usenatbib]{mnras}
\usepackage{newtxtext,newtxmath}
\usepackage[T1]{fontenc}
\usepackage{ae,aecompl}
\usepackage{epsfig}
\usepackage{amsmath, amsfonts, epsfig, xspace}
\usepackage{natbib}
\usepackage{deluxetable}
\usepackage{rotating}
\usepackage{graphicx}
\usepackage{caption}
\usepackage{longtable}
\usepackage{multicol}
\usepackage{booktabs}
\usepackage{dcolumn}
\usepackage{longtable}

\usepackage{subcaption}
\usepackage[font=small,labelfont=bf]{caption}
\def\aj{{AJ}}%
\def\araa{{ARA\&A}}%
\def\apj{{ApJ}}%
\def\apjl{{ApJ}}%
\def\apjs{{ApJS}}%
\def\aap{{A\&A}}%
\def\mnras{{MNRAS}}%
\def\pasp{{PASP}}%
\title[The alternating `changing-look' blazar OQ 334]{The alternating `changing-look' blazar OQ 334 (B2 1420+32): New observational clues to the blazar state transitions}
\author[ Krishan Chand and Gopal-Krishna]{{ \large Krishan Chand$^{1}$\thanks{E-mail: krishanchand.kc007@gmail.com (KC)} and Gopal-Krishna$^{2}$}\\\\
    $^{1}$Department of Physics and Astronomical Science, Central University of Himachal Pradesh (CUHP), Dharamshala 176215, India\\
$^{2}$UM-DAE Centre for Excellence in Basic Sciences, Vidyanagari, Mumbai 400098, India\\
}
\begin{document}
\date{Accepted ---; Received ---; in original form ---}

\pagerange{\pageref{firstpage}--\pageref{lastpage}} \pubyear{2025}

\maketitle

\label{firstpage}

\begin{abstract}
{The high-luminosity blazar OQ 334 is a leading exponent of the intriguing rare phenomenon of alternating between a flat-spectrum radio quasar (FSRQ) and a BL Lac (BLL) states. Its two optical continuum outbursts observed during the $\sim$ 1.5-year long time span, starting Jan 2018, had been shown to coincide with transition from the FSRQ to BLL state, manifested by a sharp drop in the equivalent width of MgII broad emission-line. Recently, a continuous monitoring of its blazar state, over a much longer duration (past $\sim$ 5 years) has become possible by deploying the observed $\gamma-ray$ spectral slope ($\Gamma_{\gamma}$) as a diagnostic.
%of the two blazar states. 
This opens prospects of making a much less biased and statistically more robust check on the association of optical flaring with FSRQ $\rightarrow$ BLL transition. We find that all 4 such transitions ($\Gamma_{\gamma}$ becoming < 2.0), observed during the past $\sim$ 5 years, were accompanied by an optical flare.
%lasting $\sim$ 10-15 days.and having amplitudes in the range 0.5 to 1.5-mag ($r$-band), which has been deemed in the literature as capable of swamping out the broad emission-lines in the optical spectrum of this blazar. 
While this appears consistent with the transition {to BL Lac state} happening purely due to an enhanced optical continuum (flaring) swamping out the broad emission-lines, {this simple scheme may need additional ingredients, considering the hint found for a day-like offset between the flaring and the state transition.}}

\end{abstract}

\section{Introduction}
\label{introduction}
Blazars are the most variable (non-transient) phenomenon associated with Active Galactic Nuclei (AGN). This is mainly because their radiation from millimetre to TeV bands arises substantially, often predominantly, from a Doppler-boosted relativistic jet (\citealp{Begelman1984RvMP...56..255B} and references therein). Blazars are broadly divided into two major categories—BL Lac objects (BLLs) and Flat-Spectrum Radio Quasars (FSRQs), based on the criterion of rest-frame equivalent width (EW) of the strongest broad emission-line, which is conventionally set at 5 {\AA} (rest-frame) below which a blazar is designated as BLL (e.g., \citealp{Stocke1991ApJS...76..813S}; \citealp{Stickel1991ApJ...374..431S}; \citealp{Urry1995PASP..107..803U}).
%; \citealp{Marcha1996MNRAS.281..425M}). 
An alternative classification scheme is based on the total broad-line luminosity in the units of the Eddington luminosity, with the division set at $\sim 10^{-3}L_{Edd}$ (\citealp{Ghisellini2011MNRAS.414.2674G}).
%; also, \citealp{Giommi2012MNRAS.420.2899G}; \citealp{Sbarrato2012MNRAS.421.1764S}).

According to the current AGN paradigm, the key components of their `central engine' are (i) a central supermassive black hole (SMBH), (ii) an accretion disk feeding the SMBH, (iii) a broad emission-line region (BLR) surrounding the disk and itself enveloped by a much larger `narrow-line emitting region' (NLR), and in some cases also, (iv) a `torus' of dusty material around the first three entities and (v) a bi-polar relativistic jet producing beamed nonthermal radiation, which can sometimes extend even on megaparsec scale. Interplay among these constituents, combined with their orientation relative to the line-of-sight is believed to be primarily responsible for the rich diversity of the AGN zoo (see, \citealp{Blandford2019ARA&A..57..467B} for a review). From a theoretical standpoint, concepts like `accretion disk-jet connection' (e.g., \citealp{Falcke1995A&A...293..665F}; \citealp{Marscher2002Natur.417..625M}; \citealp{Chatterjee2009ApJ...704.1689C}) have sought to probe the inter-dependences of some of the basic AGN sub-systems mentioned above. On the observational side, vital clues on the central engine have emerged from combining optical/UV spectro-polarimetry with radio VLBI observations (for reviews, see, \citealp{Antonucci1993ARA&A..31..473A, Antonucci2023Galax..11..102A}; \citealp{Urry1995PASP..107..803U}; \citealp{Blandford2019ARA&A..57..467B}) and now increasingly from the large spectroscopic and time-domain surveys revealing examples of AGN with changed classification, as manifested by radio indicators (e.g., \citealp{Nyland2020ApJ...905...74N}) and perhaps even more vividly, in the optical/UV domain (e.g., \citealp{Ricci2023NatAs...7.1282R} and references therein; \citealp{Krishan2022MNRAS.516L..18C} and references therein). In the latter category, the most vital clues are likely to come from the so-called `changing-look'  blazars which are found to transit from the FSRQ to BLL  state, or vice versa, on month-like or even shorter time scales, as manifested by the changes in their spectral properties from optical to $\gamma-rays$ (e.g., 
%\citealp{Matt2003MNRAS.342..422M}; \citealp{Bianchi2005A&A...442..185B}; \citealp{Ghisellini2011MNRAS.414.2674G}; \citealp{Marchese2012MNRAS.421.1803M}; %\citealp{Ruan2014ApJ...797...19R}; \citealp{Shappee2014ApJ...788...48S}; \citealp{Crespo2016AJ....151...32A}; \citealp{Kollatschny2020A&A...638A..91K}; %\citealp{Feng2021ApJ...909...18F}; \citealp{Foschini2022Univ....8..587F}; 
\citealp{Ricci2023NatAs...7.1282R} and references therein; \citealp{Kang2023MNRAS.525.3201K}; \citealp{Ren2024ApJ...976..124R}). 
It is clearly important to understand the physical processes underlying the blazar state transitions, also because they have the potential to challenge the canonical orientation-based AGN unification scheme (see, \citealp{Antonucci2023Galax..11..102A} and references therein). It has been proposed that the differentiation between FSRQs and BLLs is related to accretion rate changes (\citealp[e.g.,][]{Elitzur2006ApJ...648L.101E}; \citealp{Boula2019MNRAS.482L..80B}). A competing hypothesis for the FSRQ $\rightarrow$ BLL transition invokes swamping out of the broad emission-lines (`BEL swamping') by an optical continuum outburst (see Sect. \ref{OQ334}). 
%Accordingly, a re-classification from BLL to FSRQ might relate to a state change of the central engine, from a radiatively inefficient to efficient accretion disk (e.g., \citealp{Wang2002ApJ...579..554W}; \citealp{Maraschi2003ApJ...593..667M}; \citealp{Ghisellini2008MNRAS.387.1669G}; \citealp{Giommi2012MNRAS.420.2899G}; \citealp{Kondapally2024MNRAS.tmp.2502K}).

An early case that sparked the notion of `changing-look blazar' or `transition blazar' is the eponymous BL Lacertae object VRO 42.22.01 itself, whose H$\alpha$ luminosity rose several times in 1995 compared to its earlier spectroscopic observations (\citealp{Vermeulen1995ApJ...452L...5V}; \citealp{Corbett1996MNRAS.281..737C}). More such examples have been unveiled (e.g., \citealp{Stickel1991ApJ...374..431S}; \citealp{Stocke1991ApJS...76..813S}; \citealp{Ruan2014ApJ...797...19R}; \citealp{Danforth2016ApJ...832...76D}; \citealp{Ricci2023NatAs...7.1282R}, for a review). Particularly interesting are the blazars which exhibit repeated transitions, alternating between BLL and FSRQ states (e.g., \citealp{Ghisellini2013MNRAS.432L..66G}, \citealp{Ruan2014ApJ...797...19R}).
%; \citealp{Mishra2021ApJ...913..146M}; \citealp{Ren2024A&A...685A.140R}). 
The best known example, the blazar OQ 334 (J14 22 30.37+32 23 10.4, B2 1420+32) (\citealp{Mishra2021ApJ...913..146M}, hereafter, M21; \citealp{Ren2024ApJ...976..124R}, hereafter R24)
with $z$ = 0.6819 (\citealp{Hewett2010MNRAS.405.2302H}) and SMBH mass of $4\times10^8$ $M_{\odot}$ (\citealp{Brotherton2015MNRAS.454.3864B}), will be used here as a touchstone to probe this phenomenon, specifically in the context of powerful jetted AGNs. Recall that this blazar was unambiguously classified as an FSRQ during the SDSS survey (\citealp{York2000AJ....120.1579Y}; \citealp{Abdollahi2020ApJS..247...33A}), as its optical spectrum taken on MJD 53472 (2005-April-12) prominently showed broad emission-lines of MgII, H$\beta$ and H$\gamma$ (Fig. 4 of M21).

\section{The alternating blazar state of OQ 334} \label{OQ334}
We begin with a brief recapitulation of the alternating `changing-look’  pattern of this blazar, as revealed by its optical (M21) and $\gamma-ray$ spectra (R24). As mentioned above, using the SDSS spectrum taken in 2005, it was unambiguously classified as an FSRQ. Following the detection of a > 2-mag flare in December 2017 (\citealp{Stanek2017ATel11110....1S}), its optical spectroscopic follow-up was initiated  (M21), along with an extensive multi-colour optical monitoring. These light-curves (LCs) going up to May 2020, exhibited 3 optical outbursts: a flare peaking around MJD 58120 and lasting $\sim$ 1 week, and two broader outbursts peaking around MJD 58680 and MJD 58870, each lasting about a month. Monitoring by Fermi/LAT (\citealp{Atwood2009ApJ...697.1071A}) showed a conspicuous $\gamma-ray$ (0.1 - 500 GeV) flare contemporaneously to only the third optical outburst. Unfortunately, no optical spectroscopic data are available for that outburst, precluding a definitive classification of the blazar state (BLL or FSRQ) during that outburst. Spectroscopic coverage for the first two optical outbursts has been reported in M21 in the form of optical spectra taken at 8 epochs, together spanning $\sim$ 1.5 years, the last one taken on MJD 58677 (2019-July-13). The first 3 and the last 2 spectra exhibit very weak BLR signatures, signifying a BLL state, and it is these two sets of spectra (separated by $\sim$ 1.5 years) which coincide with the first two of the 3 optical outbursts mentioned above. Thus, this spectroscopic sequence revealed an FSRQ $\rightarrow$ BLL transition detected in Jan 2018 (covering 2 days), followed by a revert to the FSRQ state and then, after 416 days, to the BLL state again, this time lasting $\sim$ 12 days. According to M21, during the first FSRQ to BLL transition (MJD 58122), the equivalent width (EW) of the MgII emission-line decreased by a factor of $\sim$ 10 and then it increased by a factor of 4, restoring the FSRQ state. These authors attribute the decrease in the MgII equivalent width during the two transitions to BLL states (near MJD 58120 and MJD 58680) to the observed concurrent rise in the optical continuum contributed by the relativistic jet. The role of a Doppler-boosted jet of optical continuum in swamping out any spectral lines (including the BLR emission-lines) has been considered/invoked in several studies (e.g., \citealp{Vermeulen1995ApJ...452L...5V}; \citealp{Corbett1996MNRAS.281..737C}; \citealp{Foschini2012RAA....12..359F}; \citealp{Giommi2012MNRAS.420.2899G}; \citealp{Linford2012ApJ...757...25L}; \citealp{Ruan2014ApJ...797...19R}; \citealp{Pasham2019RNAAS...3...92P}; \citealp{Mishra2021ApJ...913..146M}; \citealp{Kang2023MNRAS.525.3201K}; \citealp{Pandey2025ApJ...978..120P}).

{At this stage, we would like to underscore certain points related to the 8 optical spectra of OQ 334 reported in M21. Although the epochs of these spectra together encompass most of the 2-year long optical light-curves of this blazar presented in M21, these spectra actually cover just a tiny fraction (8 nights) of the light-curve. Secondly, unlike the MgII broad emission-line which was clearly seen to disappear/reappear in the 8 spectra, they consistently showed at best only a marginal detection of the Balmer lines, even though these lines are quite prominently present in the SDSS spectrum (Fig. 4 of M21) taken on MJD 53472, i.e., nearly a dozen years prior to the spectroscopic campaign launched by M21. Another important point to note is that out of the 8 spectra the first 3 and the last 2 were taken near the optical continuum peaks which were observed $\sim$ 1.5 years apart, and the feeble MgII broad emission-line observed in all these 5 spectra is consistent with the `BEL swamping' hypothesis for the BLL phase (Sect. \ref{OQ334}). The remaining 3 optical spectra (together spanning $\sim$ 1.1-year) were taken between the afore-mentioned two optical outbursts, on the nights marked by a low level of optical emission, and the mutual separations of these 3 spectra range between $\sim$ 100 to 300 days. Thus, it is unclear if any transitions to BLL state (manifested by a sharply reduced prominence, or absence of the MgII broad emission-line) had occurred during the vast temporal segments that remained uncovered by optical spectroscopy. The question arising then is that while the optical spectroscopy by M21 did reveal the association of the BLL state with each of the two continuum outbursts observed (separated by $\sim$ 1.5 years), did some BLL episodes occur even during the intervening long time span of low-level optical continuum,  but they were simply missed out due to the very sparse coverage by optical spectroscopy?} 
{\it Therefore, the {point} specifically investigated in this work is whether an FSRQ to BLL transition event is always accompanied by an optical outburst?} A confirmation would mean that optical continuum flaring is an integral part of the FSRQ $\rightarrow$ BLL transition process, as also envisaged in the `BEL swamping' hypothesis. Clearly, such a test would require an essentially continuous tracking of the blazar state over several years, which however is quite infeasible by means of optical spectroscopy. We shall therefore deploy here the $\gamma-ray$ spectral slope ($\Gamma_{\gamma}$) as a tracer for monitoring the blazar state of OQ 334 continuously for several years. {As a working definition, we define an optical flare in the light-curve as a temporary brightening of the source by > 1.5-mag in $g$-band and/or $r$-band, with both rise and decay times being under $\sim$ 1 month.}
%(sect. 3) over several years and apply this diagnostic to the case of the blazar OQ 334 in the next section.

\begin{figure*}
    \includegraphics[width=1.03\textwidth,height=0.57\textheight, trim=0.1cm 0cm 0.0cm 0.0cm,clip]{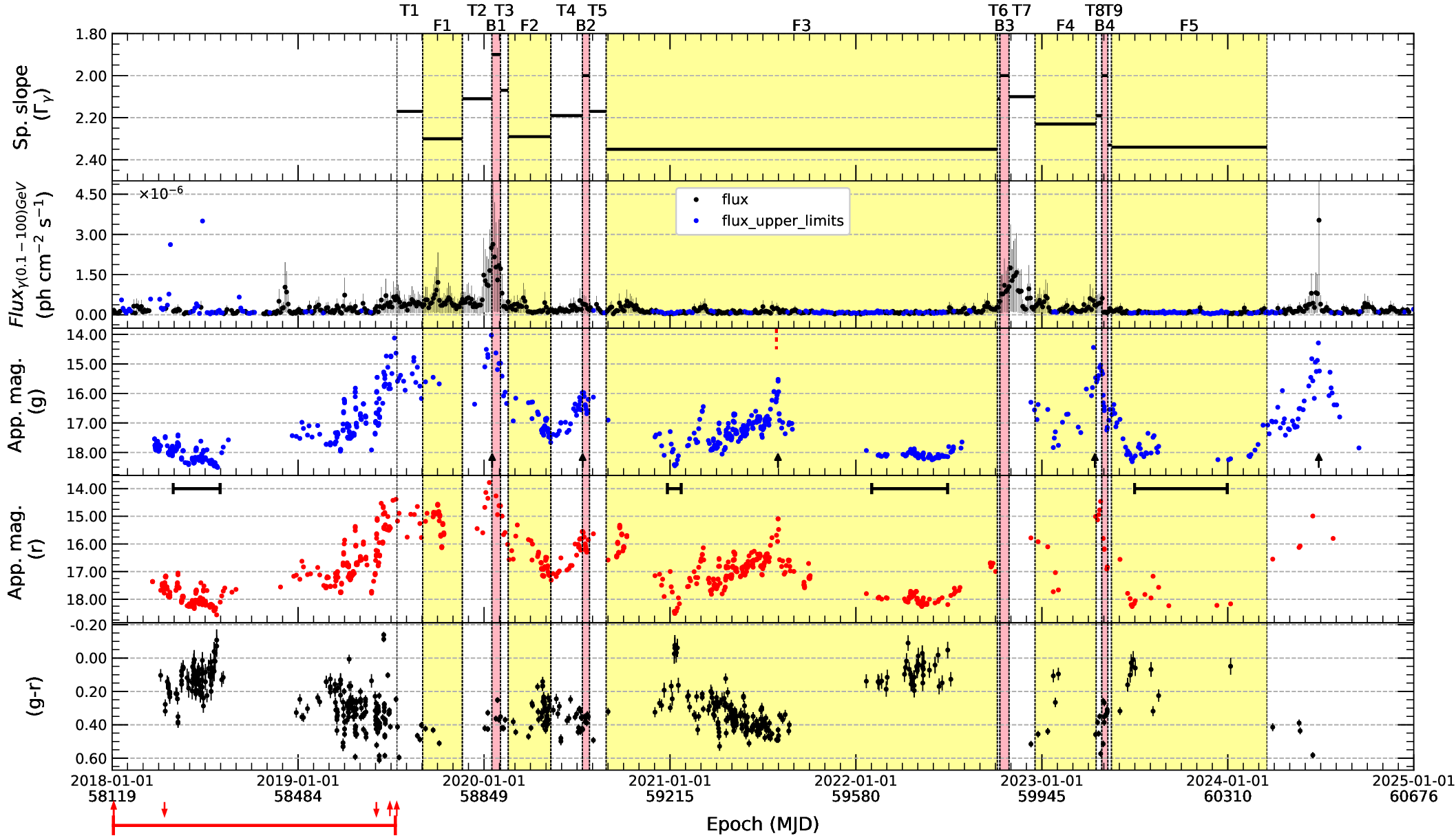}\\
  \vspace{-0.1in}
  \caption{The long-term light-curves of OQ 334 {from Fermi-LAT and ZTF surveys. The top panel shows the $\gamma-ray$ spectral slope (black horizontal bars), used to classify the state of OQ 334 (Sect. \ref{Tracking}). The second panel presents the $\gamma-ray$ light-curve, while the third and fourth panels display the ZTF light-curves in the $g$- and $r$-bands, respectively. The corresponding values of $(g-r)$ colour index are plotted in the bottom panel}. The yellow-shaded stripes show the time slots when the blazar was found to be in the FSRQ state (labelled F1, F2, F3, F4 and F5). The pink-shaded stripes correspond to the BL Lac state (labelled B1, B2, B3 and B4), while the time intervals shown with white stripes depict the state transition (labelled T1, T2, T3, T4, T5, T6, T7, T8 and T9). These classifications are based on the measured $\gamma-ray$ spectral slope (see, \citealp{Ren2024ApJ...976..124R}; Sect. \ref{Tracking}, {also the top panel}). The vertical dashed line near the centre in {the middle panel} corresponds to the epoch of the optical polarization measurement (Sect. \ref{pol}). {The upward-pointing black arrows in the middle panel indicate the peaks of the optical flares.} The red-coloured horizontal bar represents the time range covered (albeit sparsely) with optical spectroscopy (see, \citealp{Mishra2021ApJ...913..146M}). The upward-pointing {red} arrows above the bar mark the BL Lac state, while the downward-pointing {red} arrows denote the FSRQ state, as determined from the optical spectroscopy. The 4 black-coloured horizontal bars drawn in the {fourth} panel {from the top} mark the time ranges when the optical light-curves are near the base level (see Sect. \ref{pol} (c)).}
  %reported in \citet{Mishra2021ApJ...913..146M}.} 
  
\label{fig:all_dlc_part1} 
\end{figure*}

\begin{figure}
\includegraphics[width=0.51\textwidth, height=0.47\textheight, trim=7.3cm 0cm 7.0cm 0.0cm,clip]{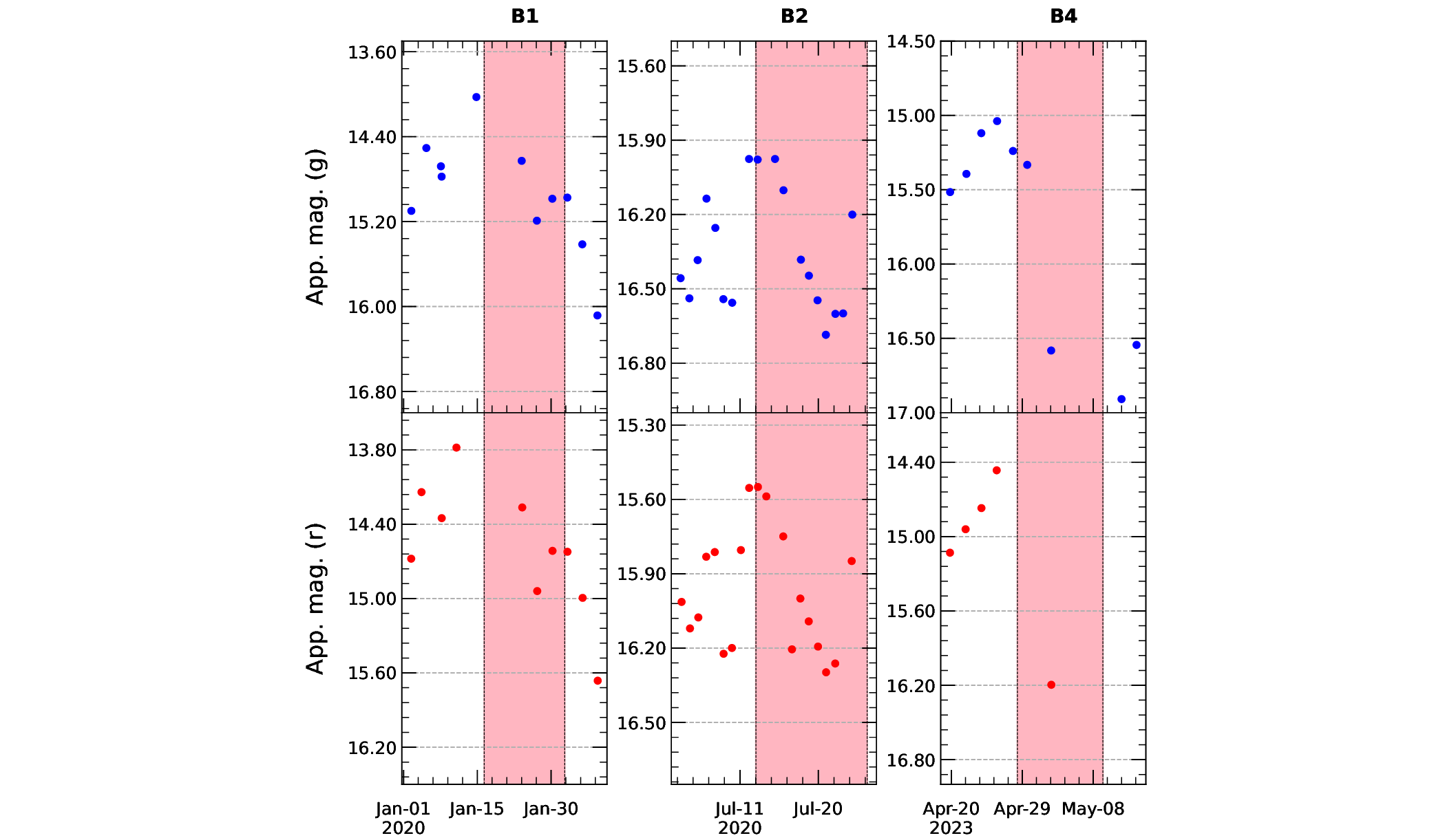}\\
  \vspace{-0.1in}
  \caption{Zoomed-in version of parts of Fig. \ref{fig:all_dlc_part1}, showing the ZTF light-curves near the transitions to the BL Lac states B1, B2 and B4.}
  \label{fig:all_dlc_part2} 
\end{figure}

%\begin{figure}
%  \centering
 % \begin{subfigure}[t]{0.257\textwidth}
  %  \includegraphics[width=\textwidth, height=0.36\textheight, trim=13.5cm 0cm 14.7cm 0.0cm, clip]{zoomed_B1_Jan25.eps}
    %\caption{B1 state}
   % \label{fig:zoomed_B1}
  %\end{subfigure}
  %\hfill
  %\begin{subfigure}[t]{0.257\textwidth}
  %  \includegraphics[width=\textwidth, height=0.36\textheight, trim=13.5cm 0cm 14.7cm 0.0cm, clip]{zoomed_B2_Jan25.eps}
    %\caption{B2 state}
   % \label{fig:zoomed_B2}
  %\end{subfigure}
  %\hfill
  %\begin{subfigure}[t]{0.257\textwidth}
   % \includegraphics[width=\textwidth, height=0.36\textheight, trim=13.5cm 0cm 14.7cm 0.0cm, clip]{zoomed_B4_Jan25.eps}
    %\caption{B4 state}
    %\label{fig:zoomed_B4}
  %\end{subfigure}
  %\caption{Zoomed version of Fig. \ref{fig:all_dlc_part1} showing $g, r, i$ light-curves near the transitions to the BL Lac states B1, B2 and B4.}
  %{\bf Increase the marker size by a factor of 2 for g, r and i-light curves. Make the shaded region lighter. make the x-tick label 2020-Jan-01, Jan-15 and Jan-30. Remove the epoch. Increase the distance between x-axis and tick label}. Increase the B1 by a factor of 2. Set Y-range for B2 from 15.00 to 17.00 (g), 15.25 to 16.75 (r) and 14.50 to 16.50 (i). Set Y-range for B4 14.50 to 17.00 (g), no change for r, no data is there.}
  %\label{fig:all_dlc_part3}
%\end{figure}

\section{Data sourcing}
\subsection{ZTF and Fermi-LAT light-curves}
{The optical light-curves in the $g$- and $r$-bands were obtained from the publicly available Zwicky Transient Facility (ZTF) survey (\citealp{Bellm2019PASP..131a8002B})\footnote{\url{https://www.ztf.caltech.edu}} which typically provides data with a cadence of $\sim$ 3 days. ZTF observes the sky in various combinations of specific field, filters and CCD quadrants, and assigns a unique observation ID to the source observed in each such configuration. Since each CCD quadrant is calibrated independently, combining light-curves (LCs) from different fields and CCD quadrants within the same filter can introduce spurious variability due to calibration differences. To mitigate this, we selected the LCs corresponding to the observation ID with the highest number of data points in each filter. Furthermore, only data recorded under favorable observing conditions as indicated by a catflag score of zero were used. \par
The $\gamma-ray$ light-curves were obtained from the Fermi-LAT survey archive, which has been providing data continuously with a sub-day cadence since its launch in 2008. We used publicly available 3-day binned light-curves covering the energy range 0.1–100 GeV, accessed through the Fermi light curve repository{\footnote{\url{https://fermi.gsfc.nasa.gov/ssc/data/access/lat/LightCurveRepository/}}}. The values of $\gamma-ray$ spectral slope of OQ 334 at  successive time intervals (Sect. \ref{Tracking}) were taken from \citet{Ren2024ApJ...976..124R} and these refer to the energy range 0.1 - 300 GeV. In addition, the optical spectroscopic and the polarization data considered here have been taken from M21 and \citet{Savchenko2021ATel14819....1S}, respectively}.\par

\section{Continuous tracking the blazar state of OQ 334 by monitoring the $\gamma-ray$ spectral slope}
\label{Tracking}

Early analyses of the Fermi-LAT observations (\citealp{Abdo2010ApJ...710.1271A}; \citealp{Ackermann2011ApJ...743..171A}) had suggested $\Gamma_{\gamma}$ = 2.2 as the dividing line between FSRQs and BLLs, albeit with a significant overlap, partly of astrophysical origin (see, also \citealp{Ghisellini2011MNRAS.414.2674G}; \citealp{Shaw2012ApJ...748...49S}). The flattening of $\Gamma_{\gamma}$ during the BLL state is probably linked to a reduction in the (external) inverse-Compton losses suffered by the jet's relativistic particles on account of shrinking/weakening of the BLR during the BLL state (e.g., \citealp{Ghisellini1998MNRAS.301..451G}; 
%Ghisellini et al. 2009; \citealp{Ghisellini2012MNRAS.425.1371G}; 
\citealp{Kang2024ApJ...962..122K}).
Regardless of the physical process involved, we adopt here the classification criterion devised by \citet{Ren2024ApJ...976..124R} which identifies $\Gamma_{\gamma}$ < 2.0 with the BLL state, $\Gamma_{\gamma}$ > 2.2 with the FSRQ state, and 2.0 < $\Gamma_{\gamma}$ < 2.2 with the transition state. This scheme is based on a multi-epoch comparison of the observed equivalent width of the MgII broad emission-line with the concurrently measured $\Gamma_{\gamma}$. Thus, applying the $\Gamma_{\gamma}$ diagnostic, these authors have mapped the distribution of the blazar states of OQ 334 across the 4.7-year time interval (MJD 58678-60387){, with MJD 58678 marking the onset of the transition state T1 (immediately following the 2-week long BL Lac state established spectroscopically by M21)}. The distribution consists of 4 BLL states, 5 FSRQ states and 9 transition states between the two (see Table 1 of \citealp{Ren2024ApJ...976..124R}, {also Fig. \ref{fig:all_dlc_part1}}). {It is noteworthy that this categorization of the blazar states {over the 4.7-year is not constrained/biased due to the usually sparse (for practical reasons) optical spectroscopic coverage, as was the case for the immediately preceding $\sim$ 1.5-year time interval covered in M21 (Sect. \ref{OQ334}).}   % Note that these spectroscopic measurements did show that during the fortnight immediately preceding the period T1 (Fig. \ref{fig:all_dlc_part1}), this source was in BL Lac state.} 
%{\bf Also, note that the distribution of the blazar state of OQ334 immediately preceding MJD 58678, spanning the 11.1-year time interval (MJD 54628–58677) was provided by Ren2024A\&A...685A.140R based on the optical spectroscopic data and the $\gamma-ray$ spectral slope ($\Gamma_{\gamma}$). The distribution consists of 2 BL Lac states, 2 FSRQ states, and 3 transition states during this time range.}  
%{\bf [Note: In \citealp{Ren2024ApJ...976..124R}, Table 1 lists a time range from MJD 58678 to 60387, during which they identified 4 BL Lac states, 5 FSRQ states, and 9 transition states. In our Fig. 1, we represent these states with pink, yellow, and white stripes, respectively for the same time range. In Fig. 3 of \citealp{Ren2024ApJ...976..124R}, the time range extends from MJD 54628 to 60387, where they report 6 BL Lac states, 7 FSRQ states, and 12 transition states. This includes data from an earlier time range (MJD 54628–58677), for which they incorporated 2 BL Lac states, 2 FSRQ states, and 3 transition states from their previous study ({Ren2024A\&A...685A.140R}).]}. 
%Note that this time span immediately preceded by (hence excludes) the 1.5-year period which M21 had covered by optical spectroscopy, albeit sparsely, leading them to identify within it two episodes of FSRQ $\rightarrow$ BLL state transitions of this blazar (Sect. \ref{OQ334}).
In essence, the above-mentioned possibility of using $\Gamma_{\gamma}$ offers an effective differentiator between the FSRQ and BLL states and thus opening up a practical way to carry out an essentially {\it continuous} monitoring of the blazar state for several years using the Fermi/LAT survey archive which contains the data with a sub-day cadence (\url{https://fermi.gsfc.nasa.gov}). The continuous, hence unbiased, mapping of the blazar states over a long time span has enabled us to now reliably address the question posed above, namely how often is a FSRQ-to-BLL transition accompanied by an optical continuum flare. Below, we investigate this point by comparing the ZTF survey optical light-curves of OQ 334 for the past $\sim$ 5 years with its blazar state monitored during the same time interval, by applying the diagnostic of $\gamma-ray$ spectral slope ($\Gamma_{\gamma}$). 
%The latter information comes from \citet{Ren2024ApJ...976..124R} who have published the run of $\Gamma_{\gamma}$ (Fig. 1), computed from the Fermi-LAT data %(\citealp{Atwood2009ApJ...697.1071A}).

\section{Results and Discussion}
\label{discussion}
Taking the blazar OQ 334 as a relatively better observed case of `alternating changing-look blazar', we shall now show that the optical flaring practically always accompanies the observed transitions from an FSRQ to BLL state, as inferred by applying the $\Gamma_{\gamma}$ < 2 criterion for the BLL state. %{\bf To identify flares in the ZTF light-curves, we define a flare as a temporary brightening of the source by a factor $\gtrsim$ 1.5-mag in either $g-$band or $r-$band, with both rise and decay times being under $\sim$ 1 month.}

\subsection{The concurrence of optical flaring with transitions to the BLL state}
%(inferred from $\Gamma_{\gamma}$)}  
\label{pol}

The evidence is gleaned here from the optical light-curves of this blazar acquired under the ZTF multi-colour survey. Fig. \ref{fig:all_dlc_part1} shows the long-term ZTF light-curves of OQ 334 in the $g$- and $r$-bands, on which the time slots occupied by the FSRQ, BLL and transition phases are marked with colour-coded stripes. These are taken from Table 1 of {R24}} 
%Ren et al. (2024b) %\citet{Ren2024ApJ...976..124R}, 
and are based on application of the $\Gamma_{\gamma}$ criterion, as mentioned above. Also shown in Fig. \ref{fig:all_dlc_part1} (bottom panel) is the run of the $g - r$ colour index, derived from the 
%ZTF $g$ and $r$-band 
light-curves plotted in the upper two panels. Only those values of colour index are displayed for which the photometric measurements in the two bands are quasi-simultaneous, to within 6 hours. The following points emerge from these plots covering the past $\sim$ 5 years:

(a) Three of the total four time slots classified as BLL (B1, B2 and B4) based on the $\Gamma_{\gamma}$ <  2.0 criterion, are clearly seen to be accompanied by an optical continuum flare seen in both optical bands.
%, whereas only the $g$ and $r$-bands measurements are available for the BLL phase B4. A good consistency is seen between the available LCs in the different bands. 
For the BLL phase B3, only $r$-band coverage is available in its vicinity. Nonetheless, a brightening of the continuum can be seen even in this sparsely sampled light-curve. Thus, evidently, all 4 time slots designated as BLL phase are accompanied by an optical continuum flare of {1.5 - 2.5-mag}. {It may be recalled that} in amplitude these flares look very similar to the two optical flares that were argued in M21 to be capable of swamping out the broad emission-lines, causing the FSRQ $\rightarrow$ BLL transitions of OQ 334. 

(b) A sharp $\sim$ 1.5-mag flare is seen in both $g$- and $r$-bands near MJD 59424 (2021-July-29) which, however, falls roughly in the middle of the 2.1-year-long time slot designated in R24 as FSRQ (shown as F3 in Fig. \ref{fig:all_dlc_part1}). Given the similarity of this optical flare (marked with an upward-pointing black arrow), both in sharpness and amplitude, to the 4 flares {shown above} to coincide with the BLL phases B1, B2, B3, B4 (Fig. \ref{fig:all_dlc_part1}), it would be worthwhile re-examining the Fermi/LAT data with a focus on just the time range occupied by this optical flare (around MJD 59424), before conclusively pronouncing its association with a FSRQ, and not with a BLL state. Indeed an independent indication for its representing BLL state comes from the observed very high degrees of optical polarization (27.5\% and 33.5\%) measured near the epoch of this flare (2021-July-28 \& 29) (\citealp{Savchenko2021ATel14819....1S}). Thus, it would be particularly interesting to re-ascertain its blazar state by estimating the value of $\Gamma_{\gamma}$ only for a narrow time window corresponding to this large optical flare. {A similar check is desirable for the marginal flare which occurred 2 months after 2021-01-01 but lacks optical polarimetric information (Fig. \ref{fig:all_dlc_part1})}.

(c) The bottom panel in Fig. \ref{fig:all_dlc_part1} displays the variation of the $g-r$ colour index 
%along with the $g-$band and  $r-$band light-curves 
of this blazar, from the beginning of the ZTF survey. 
The four horizontal bars {drawn above the $r$-band light-curve} show the time slots when the brightness appears to drop to a base level. The last 3  {of these} bars fall within the time slots identified as FSRQ state ($\Gamma_{\gamma} > 2.2$). Even the first bar is bracketed by spectroscopically confirmed FSRQ states (spectra nos. 5 and 6, see M21).
Interestingly, each of the 4 bars is accompanied by bluing of the optical continuum ($g - r$ < 0.2-mag). Very likely, this is due to domination by the (blue) accretion disk emission during these 4 base-level time slots. In contrast, all 5 sharp optical flares, shown with vertical arrows, which most probably arise due to the relativistic synchrotron jet, are accompanied by a reddening (steepening) of the optical spectrum ($g - r$ > 0.2-mag). Thus, this `alternating changing-look' blazar follows a `redder when brighter' trend which is often found to be associated with FSRQs (e.g., \citealp{Gu2006A&A...450...39G}; \citealp{Osterman2009AJ....138.1902O}; \citealp{Rani2010MNRAS.404.1992R}; \citealp{Vibhore2022MNRAS.510.1791N}).
%, and references therein) (see, however, \citealp{Ikejiri2011PASJ...63..639I}). 

To sum up, the main result emerging from the above comparison of the time sequences of optical flux and $\Gamma_{\gamma}$ for the blazar OQ 334 is that each of the 4 cases of FSRQ $\rightarrow$ BLL transition observed during the past 5 years is found to be associated with an optical flare (B1, B2, B3 and B4, see Fig. \ref{fig:all_dlc_part1}). Another flare, although seen near MJD 59424 (2021-July-29) within the time interval designated as FSRQ state (Fig. \ref{fig:all_dlc_part1}), is also likely to represent the BLL phase, considering its very high degree of polarization (see above).
 %Now, since both optical flares and $\gamma-ray$ emissions from BLLs are believed to arise predominantly from the (Doppler-boosted) relativistic jet, 
 Such temporal coincidences of the BLL phase with optical flaring is broadly in accord with the `BEL swamping' hypothesis for the FSRQ $\rightarrow$ BLL transition (Sect. \ref{introduction}). This explanation can be {further} tested/sharpened by looking for any time {offset} between the two data sequences. A preliminary test seems feasible even with the existing data, given that at least 3 out of the 4 optical flares and the corresponding BLL phases are well time-resolved at the available cadence levels of a few days, as seen in Fig. \ref{fig:all_dlc_part1}. A closer look (see the zooms in Fig. \ref{fig:all_dlc_part2}) shows that in each case the designated BLL phase is not centred on the associated optical peak, but rather it overlaps with the (post-peak) declining part of the optical flare. {However, the existing dataset permits only a crude estimate of the time lags between the 3 optical peaks and the mid-points of the corresponding BLL phases and these offsets seem to be of order of a few days (Fig. \ref{fig:all_dlc_part2}).}
 %in the g-band and 3 days in the r-band between the optical peak and the onset of the BLL phase}. 
 We {reiterate} that at present the inference of temporal offset is only a hint emerging from the available data and a definitive conclusion would need a more accurate (and denser) temporal stamping of the $\Gamma_{\gamma}$ profile and, preferably optical spectroscopic monitoring of the blazar during and around the optical flare(s), in order to track the blazar state more robustly. \par

\section{Conclusions}
\label{conclusions}
The luminous jetted AGN OQ 334 affords a powerful probe of the `changing-look' phenomenon in blazars, because of its rare property of alteranting between FSRQ and BLL states on fairly short ($\sim$ 1-year) time scales.
%(on week-to month-like time scales), 
This was established initially from optical spectroscopy which revealed that two optical continuum outbursts, separated by $\sim$ 1.5 years,  were accompanied by a transition from FSRQ to BLL state, suggesting that swamping out of the broad emission-lines (BEL) by the outbursts could have played a major role in the observed state transition. Here we have endeavoured to check whether such a state transition can occur even without an accompanying continuum outburst. To do this we have compared the $g$- and $r$-band optical light-curves of OQ 334, available in the ZTF survey archive for the past $\sim 5$ years, with the continuous monitoring of the blazar state of OQ 334, based on the monitoring of its $\gamma-ray$ spectral slope using the Fermi-LAT database. We find that each of the 4 episodes of FSRQ $\rightarrow$ BLL transition occurring during the past $\sim$ 5 years was associated with an optical continuum flare of approximately 1.5 - 2.5-mag. Thus, flares of the optical synchrotron continuum seem to be germane to the process of FSRQ $\rightarrow$ BLL transition and this indeed appears broadly consistent with the `BEL swamping' hypothesis to explain such state transitions. At the same time, we also notice a hint of potential complication to this simple scenario. This emerges from the present finding that the duration of the BLL state (as inferred from the $\Gamma_{\gamma} < 2.0$ criterion) often appears significantly offset from the optical continuum peak and it overlaps with the post-peak fading part of the optical flare. It is important to verify this temporal offset through higher cadence observations, preferably optical spectroscopy, covering the  optical flaring episodes of this leading example of an `alternating state' blazar. 

\section*{Acknowledgments}
 
 {We thank an anonymous reviewer for constructive comments.} KC acknowledges the support of SERB-DST, New Delhi, for funding under the National Post-Doctoral Fellowship Scheme through grant no. PDF/2023/004071. 
 G-K thanks Indian National Science Academy for a Senior Scientist position during which part of this work was done. The local hospitality extended to KC by IUCAA, Pune, during the preparation of this manuscript, is also thankfully acknowledged.
 This work is also based on observations obtained with the Samuel Oschin Telescope 48-inch and the 60-inch Telescope at the Palomar Observatory as part of the Zwicky Transient Facility project. ZTF is supported by the National Science Foundation under Grants No. AST-1440341 and AST-2034437 and a collaboration including current partners Caltech, IPAC, the Oskar Klein Center at Stockholm University, the University of Maryland, University of California, Berkeley, the University of Wisconsin at Milwaukee, University of Warwick, Ruhr University, Cornell University, Northwestern University and Drexel University. Operations are conducted by COO, IPAC, and UW.
\section*{Data availability}
The data used in this study are publicly available.
%The data used in this study will be shared at a reasonable request by the corresponding author.

%%%%===========================================================

%\vspace{-0.1in}
\bibliography{references}

\begin{thebibliography}{}
\makeatletter
\relax
\def\mn@urlcharsother{\let\do\@makeother \do\$\do\&\do\#\do\^\do\_\do\%\do\~}
\def\mn@doi{\begingroup\mn@urlcharsother \@ifnextchar [ {\mn@doi@} {\mn@doi@[]}}
\def\mn@doi@[#1]#2{\def\@tempa{#1}\ifx\@tempa\@empty \href {http://dx.doi.org/#2} {doi:#2}\else \href {http://dx.doi.org/#2} {#1}\fi \endgroup}
\def\mn@eprint#1#2{\mn@eprint@#1:#2::\@nil}
\def\mn@eprint@arXiv#1{\href {http://arxiv.org/abs/#1} {{\tt arXiv:#1}}}
\def\mn@eprint@dblp#1{\href {http://dblp.uni-trier.de/rec/bibtex/#1.xml} {dblp:#1}}
\def\mn@eprint@#1:#2:#3:#4\@nil{\def\@tempa {#1}\def\@tempb {#2}\def\@tempc {#3}\ifx \@tempc \@empty \let \@tempc \@tempb \let \@tempb \@tempa \fi \ifx \@tempb \@empty \def\@tempb {arXiv}\fi \@ifundefined {mn@eprint@\@tempb}{\@tempb:\@tempc}{\expandafter \expandafter \csname mn@eprint@\@tempb\endcsname \expandafter{\@tempc}}}

\bibitem[\protect\citeauthoryear{{Abdo} et~al.,}{{Abdo} et~al.}{2010}]{Abdo2010ApJ...710.1271A}
{Abdo} A.~A.,  et~al., 2010, \mn@doi [\apj] {10.1088/0004-637X/710/2/1271}, \href {https://ui.adsabs.harvard.edu/abs/2010ApJ...710.1271A} {710, 1271}

\bibitem[\protect\citeauthoryear{{Abdollahi} et~al.,}{{Abdollahi} et~al.}{2020}]{Abdollahi2020ApJS..247...33A}
{Abdollahi} S.,  et~al., 2020, \mn@doi [\apjs] {10.3847/1538-4365/ab6bcb}, \href {https://ui.adsabs.harvard.edu/abs/2020ApJS..247...33A} {247, 33}

\bibitem[\protect\citeauthoryear{{Ackermann} et~al.,}{{Ackermann} et~al.}{2011}]{Ackermann2011ApJ...743..171A}
{Ackermann} M.,  et~al., 2011, \mn@doi [\apj] {10.1088/0004-637X/743/2/171}, \href {https://ui.adsabs.harvard.edu/abs/2011ApJ...743..171A} {743, 171}

\bibitem[\protect\citeauthoryear{{Antonucci}}{{Antonucci}}{1993}]{Antonucci1993ARA&A..31..473A}
{Antonucci} R.,  1993, \mn@doi [\araa] {10.1146/annurev.aa.31.090193.002353}, \href {https://ui.adsabs.harvard.edu/abs/1993ARA&A..31..473A} {31, 473}

\bibitem[\protect\citeauthoryear{{Antonucci}}{{Antonucci}}{2023}]{Antonucci2023Galax..11..102A}
{Antonucci} R. R.~J.,  2023, \mn@doi [Galaxies] {10.3390/galaxies11050102}, \href {https://ui.adsabs.harvard.edu/abs/2023Galax..11..102A} {11, 102}

\bibitem[\protect\citeauthoryear{{Atwood} et~al.,}{{Atwood} et~al.}{2009}]{Atwood2009ApJ...697.1071A}
{Atwood} W.~B.,  et~al., 2009, \mn@doi [\apj] {10.1088/0004-637X/697/2/1071}, \href {https://ui.adsabs.harvard.edu/abs/2009ApJ...697.1071A} {697, 1071}

\bibitem[\protect\citeauthoryear{{Begelman}, {Blandford}  \& {Rees}}{{Begelman} et~al.}{1984}]{Begelman1984RvMP...56..255B}
{Begelman} M.~C.,  {Blandford} R.~D.,   {Rees} M.~J.,  1984, \mn@doi [Reviews of Modern Physics] {10.1103/RevModPhys.56.255}, \href {https://ui.adsabs.harvard.edu/abs/1984RvMP...56..255B} {56, 255}

\bibitem[\protect\citeauthoryear{{Bellm} et~al.,}{{Bellm} et~al.}{2019}]{Bellm2019PASP..131a8002B}
{Bellm} E.~C.,  et~al., 2019, \mn@doi [\pasp] {10.1088/1538-3873/aaecbe}, \href {https://ui.adsabs.harvard.edu/abs/2019PASP..131a8002B} {131, 018002}

\bibitem[\protect\citeauthoryear{{Blandford}, {Meier}  \& {Readhead}}{{Blandford} et~al.}{2019}]{Blandford2019ARA&A..57..467B}
{Blandford} R.,  {Meier} D.,   {Readhead} A.,  2019, \mn@doi [\araa] {10.1146/annurev-astro-081817-051948}, \href {https://ui.adsabs.harvard.edu/abs/2019ARA&A..57..467B} {57, 467}

\bibitem[\protect\citeauthoryear{{Boula}, {Kazanas}  \& {Mastichiadis}}{{Boula} et~al.}{2019}]{Boula2019MNRAS.482L..80B}
{Boula} S.,  {Kazanas} D.,   {Mastichiadis} A.,  2019, \mn@doi [\mnras] {10.1093/mnrasl/sly189}, \href {https://ui.adsabs.harvard.edu/abs/2019MNRAS.482L..80B} {482, L80}

\bibitem[\protect\citeauthoryear{{Brotherton}, {Singh}  \& {Runnoe}}{{Brotherton} et~al.}{2015}]{Brotherton2015MNRAS.454.3864B}
{Brotherton} M.~S.,  {Singh} V.,   {Runnoe} J.,  2015, \mn@doi [\mnras] {10.1093/mnras/stv2186}, \href {https://ui.adsabs.harvard.edu/abs/2015MNRAS.454.3864B} {454, 3864}

\bibitem[\protect\citeauthoryear{{Chand} \& {Gopal-Krishna}}{{Chand} \& {Gopal-Krishna}}{2022}]{Krishan2022MNRAS.516L..18C}
{Chand} K.,  {Gopal-Krishna} 2022, \mn@doi [\mnras] {10.1093/mnrasl/slac066}, \href {https://ui.adsabs.harvard.edu/abs/2022MNRAS.516L..18C} {516, L18}

\bibitem[\protect\citeauthoryear{{Chatterjee} et~al.,}{{Chatterjee} et~al.}{2009}]{Chatterjee2009ApJ...704.1689C}
{Chatterjee} R.,  et~al., 2009, \mn@doi [\apj] {10.1088/0004-637X/704/2/1689}, \href {https://ui.adsabs.harvard.edu/abs/2009ApJ...704.1689C} {704, 1689}

\bibitem[\protect\citeauthoryear{{Corbett}, {Robinson}, {Axon}, {Hough}, {Jeffries}, {Thurston}  \& {Young}}{{Corbett} et~al.}{1996}]{Corbett1996MNRAS.281..737C}
{Corbett} E.~A.,  {Robinson} A.,  {Axon} D.~J.,  {Hough} J.~H.,  {Jeffries} R.~D.,  {Thurston} M.~R.,   {Young} S.,  1996, \mn@doi [\mnras] {10.1093/mnras/281.3.737}, \href {https://ui.adsabs.harvard.edu/abs/1996MNRAS.281..737C} {281, 737}

\bibitem[\protect\citeauthoryear{{Danforth}, {Stocke}, {France}, {Begelman}  \& {Perlman}}{{Danforth} et~al.}{2016}]{Danforth2016ApJ...832...76D}
{Danforth} C.~W.,  {Stocke} J.~T.,  {France} K.,  {Begelman} M.~C.,   {Perlman} E.,  2016, \mn@doi [\apj] {10.3847/0004-637X/832/1/76}, \href {https://ui.adsabs.harvard.edu/abs/2016ApJ...832...76D} {832, 76}

\bibitem[\protect\citeauthoryear{{Elitzur} \& {Shlosman}}{{Elitzur} \& {Shlosman}}{2006}]{Elitzur2006ApJ...648L.101E}
{Elitzur} M.,  {Shlosman} I.,  2006, \mn@doi [\apjl] {10.1086/508158}, \href {https://ui.adsabs.harvard.edu/abs/2006ApJ...648L.101E} {648, L101}

\bibitem[\protect\citeauthoryear{{Falcke} \& {Biermann}}{{Falcke} \& {Biermann}}{1995}]{Falcke1995A&A...293..665F}
{Falcke} H.,  {Biermann} P.~L.,  1995, \mn@doi [\aap] {10.48550/arXiv.astro-ph/9411096}, \href {https://ui.adsabs.harvard.edu/abs/1995A&A...293..665F} {293, 665}

\bibitem[\protect\citeauthoryear{{Foschini}}{{Foschini}}{2012}]{Foschini2012RAA....12..359F}
{Foschini} L.,  2012, \mn@doi [Research in Astronomy and Astrophysics] {10.1088/1674-4527/12/4/001}, \href {https://ui.adsabs.harvard.edu/abs/2012RAA....12..359F} {12, 359}

\bibitem[\protect\citeauthoryear{{Ghisellini}, {Celotti}, {Fossati}, {Maraschi}  \& {Comastri}}{{Ghisellini} et~al.}{1998}]{Ghisellini1998MNRAS.301..451G}
{Ghisellini} G.,  {Celotti} A.,  {Fossati} G.,  {Maraschi} L.,   {Comastri} A.,  1998, \mn@doi [\mnras] {10.1046/j.1365-8711.1998.02032.x}, \href {https://ui.adsabs.harvard.edu/abs/1998MNRAS.301..451G} {301, 451}

\bibitem[\protect\citeauthoryear{{Ghisellini}, {Tavecchio}, {Foschini}  \& {Ghirlanda}}{{Ghisellini} et~al.}{2011}]{Ghisellini2011MNRAS.414.2674G}
{Ghisellini} G.,  {Tavecchio} F.,  {Foschini} L.,   {Ghirlanda} G.,  2011, \mn@doi [\mnras] {10.1111/j.1365-2966.2011.18578.x}, \href {https://ui.adsabs.harvard.edu/abs/2011MNRAS.414.2674G} {414, 2674}

\bibitem[\protect\citeauthoryear{{Ghisellini}, {Tavecchio}, {Foschini}, {Bonnoli}  \& {Tagliaferri}}{{Ghisellini} et~al.}{2013}]{Ghisellini2013MNRAS.432L..66G}
{Ghisellini} G.,  {Tavecchio} F.,  {Foschini} L.,  {Bonnoli} G.,   {Tagliaferri} G.,  2013, \mn@doi [\mnras] {10.1093/mnrasl/slt041}, \href {https://ui.adsabs.harvard.edu/abs/2013MNRAS.432L..66G} {432, L66}

\bibitem[\protect\citeauthoryear{{Giommi}, {Padovani}, {Polenta}, {Turriziani}, {D'Elia}  \& {Piranomonte}}{{Giommi} et~al.}{2012}]{Giommi2012MNRAS.420.2899G}
{Giommi} P.,  {Padovani} P.,  {Polenta} G.,  {Turriziani} S.,  {D'Elia} V.,   {Piranomonte} S.,  2012, \mn@doi [\mnras] {10.1111/j.1365-2966.2011.20044.x}, \href {https://ui.adsabs.harvard.edu/abs/2012MNRAS.420.2899G} {420, 2899}

\bibitem[\protect\citeauthoryear{{Gu}, {Lee}, {Pak}, {Yim}  \& {Fletcher}}{{Gu} et~al.}{2006}]{Gu2006A&A...450...39G}
{Gu} M.~F.,  {Lee} C.~U.,  {Pak} S.,  {Yim} H.~S.,   {Fletcher} A.~B.,  2006, \mn@doi [\aap] {10.1051/0004-6361:20054271}, \href {https://ui.adsabs.harvard.edu/abs/2006A&A...450...39G} {450, 39}

\bibitem[\protect\citeauthoryear{{Hewett} \& {Wild}}{{Hewett} \& {Wild}}{2010}]{Hewett2010MNRAS.405.2302H}
{Hewett} P.~C.,  {Wild} V.,  2010, \mn@doi [\mnras] {10.1111/j.1365-2966.2010.16648.x}, \href {https://ui.adsabs.harvard.edu/abs/2010MNRAS.405.2302H} {405, 2302}

\bibitem[\protect\citeauthoryear{{Kang}, {Zheng}  \& {Wu}}{{Kang} et~al.}{2023}]{Kang2023MNRAS.525.3201K}
{Kang} S.-J.,  {Zheng} Y.-G.,   {Wu} Q.,  2023, \mn@doi [\mnras] {10.1093/mnras/stad2456}, \href {https://ui.adsabs.harvard.edu/abs/2023MNRAS.525.3201K} {525, 3201}

\bibitem[\protect\citeauthoryear{{Kang}, {Lyu}, {Wu}, {Zheng}  \& {Fan}}{{Kang} et~al.}{2024}]{Kang2024ApJ...962..122K}
{Kang} S.-J.,  {Lyu} B.,  {Wu} Q.,  {Zheng} Y.-G.,   {Fan} J.,  2024, \mn@doi [\apj] {10.3847/1538-4357/ad0fdf}, \href {https://ui.adsabs.harvard.edu/abs/2024ApJ...962..122K} {962, 122}

\bibitem[\protect\citeauthoryear{{Linford}, {Taylor}  \& {Schinzel}}{{Linford} et~al.}{2012}]{Linford2012ApJ...757...25L}
{Linford} J.~D.,  {Taylor} G.~B.,   {Schinzel} F.~K.,  2012, \mn@doi [\apj] {10.1088/0004-637X/757/1/25}, \href {https://ui.adsabs.harvard.edu/abs/2012ApJ...757...25L} {757, 25}

\bibitem[\protect\citeauthoryear{{Marscher}, {Jorstad}, {G{\'o}mez}, {Aller}, {Ter{\"a}sranta}, {Lister}  \& {Stirling}}{{Marscher} et~al.}{2002}]{Marscher2002Natur.417..625M}
{Marscher} A.~P.,  {Jorstad} S.~G.,  {G{\'o}mez} J.-L.,  {Aller} M.~F.,  {Ter{\"a}sranta} H.,  {Lister} M.~L.,   {Stirling} A.~M.,  2002, \mn@doi [\nat] {10.1038/nature00772}, \href {https://ui.adsabs.harvard.edu/abs/2002Natur.417..625M} {417, 625}

\bibitem[\protect\citeauthoryear{{Mishra} et~al.,}{{Mishra} et~al.}{2021}]{Mishra2021ApJ...913..146M}
{Mishra} H.~D.,  et~al., 2021, \mn@doi [\apj] {10.3847/1538-4357/abf63d}, \href {https://ui.adsabs.harvard.edu/abs/2021ApJ...913..146M} {913, 146}

\bibitem[\protect\citeauthoryear{{Negi}, {Joshi}, {Chand}, {Chand}, {Wiita}, {Ho}  \& {Singh}}{{Negi} et~al.}{2022}]{Vibhore2022MNRAS.510.1791N}
{Negi} V.,  {Joshi} R.,  {Chand} K.,  {Chand} H.,  {Wiita} P.,  {Ho} L.~C.,   {Singh} R.~S.,  2022, \mn@doi [\mnras] {10.1093/mnras/stab3591}, \href {https://ui.adsabs.harvard.edu/abs/2022MNRAS.510.1791N} {510, 1791}

\bibitem[\protect\citeauthoryear{{Nyland} et~al.,}{{Nyland} et~al.}{2020}]{Nyland2020ApJ...905...74N}
{Nyland} K.,  et~al., 2020, \mn@doi [\apj] {10.3847/1538-4357/abc341}, \href {https://ui.adsabs.harvard.edu/abs/2020ApJ...905...74N} {905, 74}

\bibitem[\protect\citeauthoryear{{Osterman Meyer}, {Miller}, {Marshall}, {Ryle}, {Aller}, {Aller}  \& {Balonek}}{{Osterman Meyer} et~al.}{2009}]{Osterman2009AJ....138.1902O}
{Osterman Meyer} A.,  {Miller} H.~R.,  {Marshall} K.,  {Ryle} W.~T.,  {Aller} H.,  {Aller} M.,   {Balonek} T.,  2009, \mn@doi [\aj] {10.1088/0004-6256/138/6/1902}, \href {https://ui.adsabs.harvard.edu/abs/2009AJ....138.1902O} {138, 1902}

\bibitem[\protect\citeauthoryear{{Pandey} et~al.,}{{Pandey} et~al.}{2025}]{Pandey2025ApJ...978..120P}
{Pandey} A.,  et~al., 2025, \mn@doi [\apj] {10.3847/1538-4357/ad9b7c}, \href {https://ui.adsabs.harvard.edu/abs/2025ApJ...978..120P} {978, 120}

\bibitem[\protect\citeauthoryear{{Pasham} \& {Wevers}}{{Pasham} \& {Wevers}}{2019}]{Pasham2019RNAAS...3...92P}
{Pasham} D.~R.,  {Wevers} T.,  2019, \mn@doi [Research Notes of the American Astronomical Society] {10.3847/2515-5172/ab304a}, \href {https://ui.adsabs.harvard.edu/abs/2019RNAAS...3...92P} {3, 92}

\bibitem[\protect\citeauthoryear{{Rani} et~al.,}{{Rani} et~al.}{2010}]{Rani2010MNRAS.404.1992R}
{Rani} B.,  et~al., 2010, \mn@doi [\mnras] {10.1111/j.1365-2966.2010.16419.x}, \href {https://ui.adsabs.harvard.edu/abs/2010MNRAS.404.1992R} {404, 1992}

\bibitem[\protect\citeauthoryear{{Ren}, {Zhou}, {Zheng}  \& {Kang}}{{Ren} et~al.}{2024}]{Ren2024ApJ...976..124R}
{Ren} S.~S.,  {Zhou} R.~X.,  {Zheng} Y.~G.,   {Kang} S.~J.,  2024, \mn@doi [\apj] {10.3847/1538-4357/ad83ce}, \href {https://ui.adsabs.harvard.edu/abs/2024ApJ...976..124R} {976, 124}

\bibitem[\protect\citeauthoryear{{Ricci} \& {Trakhtenbrot}}{{Ricci} \& {Trakhtenbrot}}{2023}]{Ricci2023NatAs...7.1282R}
{Ricci} C.,  {Trakhtenbrot} B.,  2023, \mn@doi [Nature Astronomy] {10.1038/s41550-023-02108-4}, \href {https://ui.adsabs.harvard.edu/abs/2023NatAs...7.1282R} {7, 1282}

\bibitem[\protect\citeauthoryear{{Ruan}, {Anderson}, {Plotkin}, {Brandt}, {Burnett}, {Myers}  \& {Schneider}}{{Ruan} et~al.}{2014}]{Ruan2014ApJ...797...19R}
{Ruan} J.~J.,  {Anderson} S.~F.,  {Plotkin} R.~M.,  {Brandt} W.~N.,  {Burnett} T.~H.,  {Myers} A.~D.,   {Schneider} D.~P.,  2014, \mn@doi [\apj] {10.1088/0004-637X/797/1/19}, \href {https://ui.adsabs.harvard.edu/abs/2014ApJ...797...19R} {797, 19}

\bibitem[\protect\citeauthoryear{{Savchenko}, {Chazov}, {Nazarova}, {Shishkina}  \& {Larionova}}{{Savchenko} et~al.}{2021}]{Savchenko2021ATel14819....1S}
{Savchenko} S.~S.,  {Chazov} M.~I.,  {Nazarova} A.~E.,  {Shishkina} E.~V.,   {Larionova} E.~G.,  2021, The Astronomer's Telegram, \href {https://ui.adsabs.harvard.edu/abs/2021ATel14819....1S} {14819, 1}

\bibitem[\protect\citeauthoryear{{Shaw} et~al.,}{{Shaw} et~al.}{2012}]{Shaw2012ApJ...748...49S}
{Shaw} M.~S.,  et~al., 2012, \mn@doi [\apj] {10.1088/0004-637X/748/1/49}, \href {https://ui.adsabs.harvard.edu/abs/2012ApJ...748...49S} {748, 49}

\bibitem[\protect\citeauthoryear{{Stanek}, {Kochanek}, {Thompson}, {Shappee}, {Holoien}, {Prieto}, {Dong}  \& {Dai}}{{Stanek} et~al.}{2017}]{Stanek2017ATel11110....1S}
{Stanek} K.~Z.,  {Kochanek} C.~S.,  {Thompson} T.~A.,  {Shappee} B.~J.,  {Holoien} T.~W.~S.,  {Prieto} J.~L.,  {Dong} S.,   {Dai} X.,  2017, The Astronomer's Telegram, \href {https://ui.adsabs.harvard.edu/abs/2017ATel11110....1S} {11110, 1}

\bibitem[\protect\citeauthoryear{{Stickel}, {Padovani}, {Urry}, {Fried}  \& {Kuehr}}{{Stickel} et~al.}{1991}]{Stickel1991ApJ...374..431S}
{Stickel} M.,  {Padovani} P.,  {Urry} C.~M.,  {Fried} J.~W.,   {Kuehr} H.,  1991, \mn@doi [\apj] {10.1086/170133}, \href {http://adsabs.harvard.edu/abs/1991ApJ...374..431S} {374, 431}

\bibitem[\protect\citeauthoryear{{Stocke}, {Morris}, {Gioia}, {Maccacaro}, {Schild}, {Wolter}, {Fleming}  \& {Henry}}{{Stocke} et~al.}{1991}]{Stocke1991ApJS...76..813S}
{Stocke} J.~T.,  {Morris} S.~L.,  {Gioia} I.~M.,  {Maccacaro} T.,  {Schild} R.,  {Wolter} A.,  {Fleming} T.~A.,   {Henry} J.~P.,  1991, \mn@doi [\apjs] {10.1086/191582}, \href {https://ui.adsabs.harvard.edu/abs/1991ApJS...76..813S} {76, 813}

\bibitem[\protect\citeauthoryear{{Urry} \& {Padovani}}{{Urry} \& {Padovani}}{1995}]{Urry1995PASP..107..803U}
{Urry} C.~M.,  {Padovani} P.,  1995, \mn@doi [\pasp] {10.1086/133630}, \href {http://adsabs.harvard.edu/abs/1995PASP..107..803U} {107, 803}

\bibitem[\protect\citeauthoryear{{Vermeulen}, {Ogle}, {Tran}, {Browne}, {Cohen}, {Readhead}, {Taylor}  \& {Goodrich}}{{Vermeulen} et~al.}{1995}]{Vermeulen1995ApJ...452L...5V}
{Vermeulen} R.~C.,  {Ogle} P.~M.,  {Tran} H.~D.,  {Browne} I.~W.~A.,  {Cohen} M.~H.,  {Readhead} A.~C.~S.,  {Taylor} G.~B.,   {Goodrich} R.~W.,  1995, \mn@doi [\apjl] {10.1086/309716}, \href {https://ui.adsabs.harvard.edu/abs/1995ApJ...452L...5V} {452, L5}

\bibitem[\protect\citeauthoryear{{York} et~al.,}{{York} et~al.}{2000}]{York2000AJ....120.1579Y}
{York} D.~G.,  et~al., 2000, \mn@doi [\aj] {10.1086/301513}, \href {http://adsabs.harvard.edu/abs/2000AJ....120.1579Y} {120, 1579}

\makeatother
\end{thebibliography}
\label{lastpage}
\end{document}